\documentclass[12pt]{iopart}
\usepackage{graphicx}

\begin{document}

\title[Stability Conditions for One Photon Loop Processes]
{Static and Dynamic Casimir Effect Instabilities}

\author{Y. N. Srivastava\dag\ddag , A. Widom\ddag ,\\
S. Sivasubramanian\S , M. Pradeep Ganesh\ddag\ }

\address{\dag\ Physics Department and \& INFN,
University of Perugia, Perugia IT}
\address{\ddag\ Physics Department,
Northeastern University, Boston MA USA}
\address{\S Department of Electrical Engineering,
Northeastern University, Boston MA USA}

\begin{abstract}
The static Casimir effect concerns quantum electrodynamic
induced Lamb shifts in the mode frequencies and thermal free energies
of condensed matter systems. Sometimes, the condensed matter
constitutes the boundaries of a vacuum region. The static frequency shift
effects have been calculated in the one photon loop perturbation theory
approximation. The dynamic Casimir effect concerns two photon radiation
processes arising from time dependent frequency modulations again computed
in the one photon loop approximation. Under certain conditions the one photon
loop computation may become unstable and higher order terms must be invoked
to achieve stable solutions. This stability calculation is discussed for
a simple example dynamical Casimir effect system.
\end{abstract}



\section{Introduction}

The Casimir effect concerns the Lamb shifts in the frequency of radiation
modes due to the interaction between photon modes and electrical currents.
The photon mode Lagrangian is discussed in Sec.\ref{LCMD}. Mode
frequency shifts induce changes in the free energy which in the zero
temperature limit\cite{Casimir:1948,Casimir::1948} reduce to changes in the zero point
energy\cite{Bordag:2001,Milton:2001}
\begin{math} \delta E_0=(\hbar /2)\sum_{a} \delta \Omega_a  \end{math}.
The Lamb frequency shifts are usually small and can be understood from a
perturbation theory viewpoint.
Such damping is discussed in Sec.\ref{OCD}. The conventional Casimir effect
theory thereby considers Feynman diagram corrections to the free energy
containing one photon loop\cite{Dzyaloshinski:1960,Dzyaloshinski:1961}.
In Sec.\ref{TS} it is shown how a one loop instability can
arise if the coupling between a photon oscillation mode
and the surrounding currents is too strong.

If the damping functions and frequency shifts are also oscillating
functions of time, then (over and above single photon absorption and
emission processes) there is the absorption and emission of
{\em photon pairs}\cite{Dodonov:1998}. The photon pair processes constitute a
dynamical Casimir effect\cite{Dalvit:1999,Dodonov:1999}. Frequency modulations
tend to heat up the cavity. In Sec.\ref{DCE}, the noise temperature description
is discussed. In Sec.\ref{PFM}, the heating of a cavity mode by periodic frequency
modulation is explored. In an unstable regime, the temperature of (say)
a microwave cavity mode grows {\em exponentially}. The implied
purely theoretical {\em microwave oven} would be much more hot than
that which could be observed in experimental reality. Nonlinear
higher loop photon processes producing dynamic microwave intensity
stability are discussed in Sec.\ref{MS}.

\section{Lagrangian Circuit Mode Description \label{LCMD}}

Our purpose in this section is to provide a Lagrangian description
of a single microwave cavity mode which follows from the action
principle formulation of electrodynamics\cite{Widom:1987}. For this purpose
we employ the Coulomb gauge, \begin{math} div{\bf A}_{mode}=0  \end{math},
for the vector potential. The vector potential representing the
cavity mode may be written
\begin{equation}
{\bf A}_{mode}({\bf r},t)=\Phi(t){\bf K}({\bf r}).
\label{ML1}
\end{equation}
The mode electromagnetic fields are then given by
\begin{eqnarray}
{\bf E}_{mode}({\bf r},t)
=-\frac{1}{c}\left[\frac{{\bf A}_{mode}({\bf r},t)}{\partial t}\right]
=-\frac{\dot{\Phi }(t)}{c}{\bf K}({\bf r}),
\nonumber \\
{\bf B}_{mode}({\bf r},t)
= curl{\bf A}_{mode}({\bf r},t)=\Phi(t)\ curl{\bf K}({\bf r}).
\label{ML2}
\end{eqnarray}
The Lagrangian
\begin{equation}
L_{field}=\frac{1}{8\pi }\int_{cavity}
\left[\left|{\bf E}_{mode}({\bf r},t)\right|^2-
\left|{\bf B}_{mode}({\bf r},t)\right|^2\right]d^3{\bf r}
\label{ML3}
\end{equation}
describes the mode in terms of a simple oscillator circuit.
The capacitance \begin{math} C \end{math} and inductance
\begin{math} \Lambda \end{math} of the circuit are defined,
respectively, by
\begin{eqnarray}
C=\frac{1}{4\pi }\int_{cavity} \left|{\bf K}({\bf r})\right|^2 d^3{\bf r},
\nonumber \\
\frac{1}{\Lambda }=
\frac{1}{4\pi }\int_{cavity} \left|curl{\bf K}({\bf r})\right|^2 d^3{\bf r}.
\label{ML4}
\end{eqnarray}
The circuit electromagnetic field Lagrangian follows from
Eqs.(\ref{ML2}), (\ref{ML3}) and (\ref{ML4}). It is of the
simple \begin{math} \Lambda C \end{math} oscillator form
\begin{equation}
L_{field}(\dot{\Phi },\Phi )=\frac{C}{2c^2}\dot{\Phi }^2
-\frac{1}{2\Lambda }\Phi^2,
\label{ML5}
\end{equation}
wherein the bare circuit frequency obeys
\begin{equation}
\Omega_\infty ^2=\frac{c^2}{\Lambda C}\ .
\label{ML6}
\end{equation}
The interactions between cavity wall currents and an electromagnetic
mode are conventionally described by
\begin{eqnarray}
L_{int}=\frac{1}{c}\int {\bf J}
\cdot {\bf A}_{mode }d^3{\bf r},
\nonumber \\
L_{int}= \frac{1}{c}I\Phi ,
\nonumber \\
I(t) = \int {\bf J}({\bf r},t)\cdot {\bf K}({\bf r})d^3{\bf r},
\label{ML7}
\end{eqnarray}
where the current \begin{math} I \end{math} drives the oscillator circuit.

In total, the circuit mode Lagrangian follows from
Eqs.(\ref{ML5}) and (\ref{ML7}) as
\begin{equation}
L=\frac{C}{2c^2}\dot{\Phi }^2
-\frac{1}{2\Lambda }\Phi^2+\frac{1}{c}I\Phi +L^\prime
\label{ML8}
\end{equation}
wherein \begin{math} L^\prime  \end{math} describes all of the
other degrees of freedom which couple into the mode coordinate.
Maxwell's equations for a single microwave mode then takes the form
\begin{eqnarray}
\frac{d}{dt}\left(\frac{\partial L}{\partial \dot{\Phi}}\right)
=\left(\frac{\partial L}{\partial \Phi}\right),
\nonumber \\
C\left(\ddot{\Phi }+\Omega_\infty ^2\Phi \right)=cI.
\label{ML9}
\end{eqnarray}
The damping of the oscillator will first be discussed from a classical
electrical engineering viewpoint and only later from a fully quantum
electrodynamic viewpoint.

\section{Oscillator Circuit Damping \label{OCD}}

From an electrical engineering viewpoint, let us consider a small external
current source \begin{math} \delta I_{ext}  \end{math} which drives the mode
coordinate \begin{math} \delta \Phi \end{math}. Eq.(\ref{ML9}) now reads
\begin{equation}
\frac{C}{c^2}\delta \ddot{\Phi }+\frac{1}{\Lambda }\delta \Phi =
\frac{1}{c}\delta I=\frac{1}{c}\left(\delta I_{ext}+\delta I_{ind}\right)
\label{CD1}
\end{equation}
were \begin{math} \delta I_{ind}  \end{math} is the current induced by the
coordinate response \begin{math} \delta \Phi \end{math}. In the complex
frequency \begin{math} \zeta \end{math} domain we have in (the upper half
\begin{math} {\Im }m\ \zeta >0  \end{math} plane)
\begin{eqnarray}
\delta I_{ext}(t) = {\Re }e\left\{\delta I_{ext,\zeta}e^{-i\zeta t}\right\}
\nonumber \\
\delta \Phi (t) = {\Re }e\left\{\delta I_{ext,\zeta}
{\cal D}(\zeta )e^{-i\zeta t}\right\}.
\label{CD2}
\end{eqnarray}
The induced current is determined by the ``surface admittance''
\begin{math} Y(\zeta ) \end{math} of the cavity walls;
In detail
\begin{eqnarray}
\delta I_{ind}(t) = -\frac{1}{c}\int_0^\infty
{\cal G}(t^\prime )\delta \dot{\Phi }(t-t^\prime ) dt^\prime ,
\nonumber \\
Y(\zeta ) = \int_0^\infty e^{i\zeta t}{\cal G}(t)dt,
\label{CD3}
\end{eqnarray}
so that
\begin{eqnarray}
\left\{-\frac{C}{c^2}\zeta ^2+\frac{1}{\Lambda }
-\frac{i\zeta }{c^2}Y(\zeta )\right\}{\cal D}(\zeta )
=\frac{1}{c}\ ,
\nonumber \\
-i\zeta \varepsilon (\zeta )C=-i\zeta C+Y(\zeta ),
\label{CD4}
\end{eqnarray}
wherein the effective frequency dependent capacitance
\begin{math} \varepsilon (\zeta )C  \end{math}
determines the mode dielectric
response function \begin{math} \varepsilon (\zeta ) \end{math}.
The retarded propagator for the mode in the frequency domain obeys
\begin{equation}
{\cal D}(\zeta )=\frac{\Lambda }{c}
\left[\frac{\Omega_\infty ^2}{\Omega_\infty ^2-\zeta ^2-\Pi(\zeta )}\right]
\label{CD5}
\end{equation}
wherein the ``self energy'' \begin{math} \Pi(\zeta ) \end{math} is
determined by the induced current admittance via
\begin{equation}
{\Pi (\zeta )}=\frac{i\zeta Y(\zeta )}{C}\ .
\label{CD6}
\end{equation}
The self energy describes both frequency shift and damping properties of the
mode.

Causality dictates that all engineering response functions obey analytic
dispersion relations (\begin{math} {\Im }m\ \zeta >0 \end{math}) of the form
\begin{eqnarray}
{\cal D}(\zeta )=\frac{2}{\pi }\int_0^\infty
\frac{\omega {\Im }m{\cal D}(\omega+i0^+)d\omega }{\omega^2 -\zeta ^2}\ ,
\nonumber \\
\Pi (\zeta )=\frac{2}{\pi }\int_0^\infty
\frac{\omega {\Im }m\Pi (\omega+i0^+)d\omega }{\omega^2 -\zeta ^2}\ .
\label{CD7}
\end{eqnarray}
The damping rate for the oscillation is determined by
\begin{equation}
{\Im }m\Pi (\omega+i0^+)=\omega {\Re}e\Gamma (\omega +i0^+)
=\frac{\omega {\Re}e Y(\omega +i0^+)}{C}\ .
\label{CD8}
\end{equation}
The shifted frequency,
\begin{equation}
\Omega_0^2=\Omega_\infty ^2-\Pi(0),
\label{CD9}
\end{equation}
obeys the dispersion relation sum rule
The shifted frequency is related to the damping rate via the sum rule
\begin{equation}
\Omega_\infty ^2=\Omega_0^2+\frac{2}{\pi }
\int_0^\infty {\Re}e\Gamma (\omega +i0^+)d\omega ,
\label{CD10}
\end{equation}
which follows from Eqs.(\ref{CD6}) - (\ref{CD9}).
Finally, the the quality factor \begin{math} Q \end{math} for the mode
frequency \begin{math} \Omega_0 \end{math} is well defined as
\begin{equation}
\frac{\Omega_0}{Q}={\Re}e\Gamma (\Omega_0 +i0^+)
\label{CD11}
\end{equation}
if and only if the mode is under damped by a large margin; e.g.
\begin{math} Q>>1  \end{math}. On the other hand,
if the damping is sufficiently strong, then the mode can go unstable.
Let us consider this physical effect in more detail.

\section{Thermodynamic Stability \label{TS}}

If the mode where uncoupled to the damping current, then then the free energy
of the oscillator would be
\begin{eqnarray}
f_\infty (T)= -k_BT
\ln \left[\sum_{N=0}^\infty e^{-(N+1/2)\hbar \Omega_\infty /k_BT}\right],
\nonumber \\
f_\infty (T)=k_BT
\ln \left[\sinh \left(\frac{\hbar \Omega_\infty}{2k_BT}\right)\right].
\label{TS1}
\end{eqnarray}
The damping effects give rise to Lamb shifted frequencies and a Casimir-Lifshitz
renormalization in the free energy; It is
\begin{eqnarray}
f(T) = f_\infty (T)+f_1(T),
\nonumber \\
f_1(T) = \left(\frac{k_BT}{2}\right)\sum_{n=-\infty}^\infty
\ln\left[1-\left(\frac{\Pi (i|\omega_n|)}{\Omega_\infty^2+\omega_n^2}\right)\right],
\nonumber \\
\hbar \omega_n = 2\pi nk_BT .
\nonumber \\
\Pi (i|\omega_n|) = \frac{2}{\pi }\int_0^\infty
\frac{\omega {\Im }m\Pi (\omega+i0^+)d\omega }{\omega^2 +\omega_n^2}\ .
\label{TS2}
\end{eqnarray}
A sufficient condition for the validity of Eqs.(\ref{TS2}) is that the mode oscillator
obeys a linear equation of motion. From Eqs.(\ref{TS1}) and (\ref{TS2}) we deduce the
following thermodynamic stability\cite{Widom:2004}
\medskip
\par \noindent
{\bf Theorem 1:} {\it The Casimir free energy shift of an oscillator mode is stable if
and only if \begin{math} \Pi (0)<\Omega_\infty ^2 \end{math}. If
\begin{math} \Pi (0)>\Omega_\infty ^2 \end{math}, then the one loop
free energy in {Eq.{\rm (\ref{TS2})}} becomes complex yielding finite lifetime effects.}
\medskip
\par \noindent
Thermodynamic stability can be restored if the one goes beyond the one loop
approximation in the effective Lagrangian, e.g. the oscillator can shift its
minimum form zero to \begin{math} \Phi_0  \end{math}. For such a thermodynamic
instability in which \begin{math} \omega_0^2=\Pi(0)-\Omega_\infty ^2>0 \end{math},
the effective Lagrangian may be taken as
\begin{equation}
L_{effective}=\frac{C}{2c^2}\dot{\Phi }^2+
\frac{C\omega_0^2}{4\Phi_0^2 c^2}\left(\Phi^2-\Phi_0^2\right)^2.
\label{TS3}
\end{equation}
The stability is restored via a stabilizing term representing four photon
interactions. Such a Lagrangian can appear for modes whose surrounding
walls are at least in part ferromagnetic.

A high quality photon oscillator mode is only weakly damped so that
the one loop perturbation approximation is virtually exact. On the
other hand {\em dynamical instabilities} may still require higher order
photon interaction terms to understand the ultimate stabilities in
laboratory systems.

\section{Dynamical Casimir Effects \label{DCE}}

Suppose that the dielectric response function
\begin{math} \varepsilon (\zeta ) \end{math} of the mode
in Eq.(\ref{CD4}) is made to vary time; i.e.
\begin{equation}
\varepsilon (\zeta )\ \Rightarrow\ \varepsilon(\zeta ,t)
\ \ {\rm equivalently}
\ \ \Pi (\zeta )\ \Rightarrow\ \Pi(\zeta ,t).
\label{DCE1}
\end{equation}
If the resulting differential equation for the
\begin{math} \Phi =\Re e \{\phi \}\end{math} signal obeys
to a sufficient degree of accuracy
\begin{eqnarray}
\ddot{\phi}(t)+\Omega^2 (t)\phi (t)=0,
\nonumber \\
\Omega (t\to \pm \infty)=\Omega_0,
\label{DCE2}
\end{eqnarray}
then there exists a solution of the form
\begin{eqnarray}
\phi (t\to \infty )=e^{i\Omega_0 t}+\rho e^{-i\Omega_0 t},
\nonumber \\
\phi (t\to -\infty )=\sigma e^{i\Omega_0 t},
\nonumber \\
|\rho |^2+|\sigma |^2=1.
\label{DCE3}
\end{eqnarray}
From a quantum mechanical viewpoint, the time variation
\begin{math} e^{i\Omega_0 t} \end{math} may represent a
photon moving backward in time and
\begin{math} e^{-i\Omega_0 t} \end{math}
may represent photon moving forward in time. In Eq.(\ref{DCE3}),
the reflection amplitude for a photon moving backward in time
to bounce forward in time is given by
\begin{math} \rho \end{math}. A backward in time moving photon
reflected forward in time appears in the laboratory to be a pair
of photons being created.

\begin{figure}[bp]
\scalebox {0.8}{\includegraphics{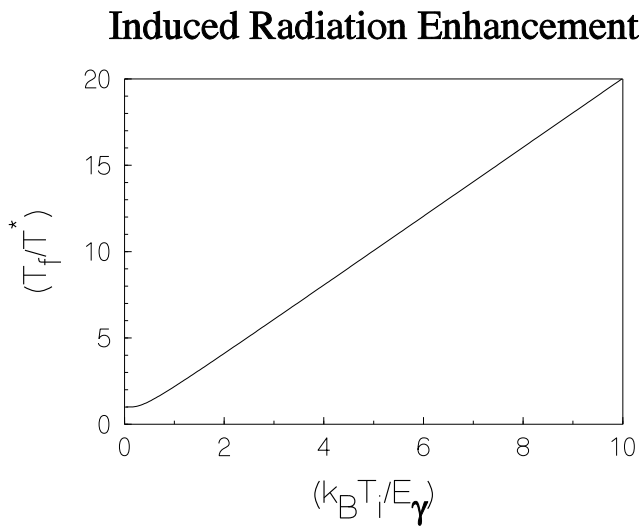}}
\caption{If $T_i$ represents the initial cavity mode temperature and $T^*$
represents the noise temperature of the pair radiated photons, then the final
temperature $T_f$ of the of the cavity mode is enhanced (over
and above $T^*)$ via the initial photon population. The resulting radiation
enhancement is plotted for photons with energy $E_\gamma =\hbar \Omega_0$.}
\label{Fig1}
\end{figure}

The probability of such a photon pair creation event defines
a {\em photon pair creation noise temperature} \begin{math} T^*  \end{math}
induced by the time varying frequency via
\begin{equation}
R=|\rho |^2=e^{-\hbar \Omega_0/k_BT^*}.
\label{DCE4}
\end{equation}
The mean number \begin{math} \bar{N}  \end{math} of photons which
would be radiated from the vacuum by a time varying frequency
modulation \begin{math} \Omega (t) \end{math} obeys a formal Planck
law
\begin{equation}
\bar{N}=\frac{R}{1-R}=\frac{1}{e^{\hbar \Omega_0/k_BT^*}-1}\ .
\label{DCE5}
\end{equation}
Suppose (for example) that a microwave cavity is initially in thermal
equilibrium at temperature \begin{math} T_i  \end{math}. The mean
number of initial microwave photons in a given normal mode is then given by
\begin{equation}
N_i=\frac{1}{e^{\hbar \Omega_0/k_BT_i}-1}\ .
\label{DCE6}
\end{equation}
After a sequence of frequency modulation pulses the mean number of final
photons in the cavity mode is
\begin{equation}
N_f=(2\bar{N}+1)N_i+\bar{N}=
N_i\coth\left(\frac{\hbar \Omega_0}{2k_BT^*}\right)
+\frac{1}{e^{\hbar \Omega_0/k_BT^*}-1}\ .
\label{DCE7}
\end{equation}
Note that the existence of an {\em initial} number of photons
\begin{math} N_i \end{math} in the cavity mode makes larger the
final number of of photons
\begin{equation}
N_f=\frac{1}{e^{\hbar \Omega_0/k_BT_f}-1}
\label{DCE6f}
\end{equation}
via the {\em induced} radiation of additional
photon pairs. If the microwave frequency large margin inequality
\begin{equation}
\hbar \Omega_0\ll k_BT^*
\label{DCE8}
\end{equation}
holds true, then Eqs.(\ref{DCE5}) - (\ref{DCE8}) imply an
approximate law for the {\em final} cavity mode noise temperature
is given by
\begin{equation}
T_f  \approx T^* \coth\left(\frac{\hbar \Omega_0}{2k_BT_i}\right).
\label{DCE9}
\end{equation}
The resulting enhancement \begin{math} (T_f/T^*)  \end{math} is
plotted in Fig.\ref{Fig1}. The dynamical Casimir effect for frequency
modulation pulses is thereby described in terms of the amount of
heat that raises the temperature \begin{math} T_i\to T_f  \end{math}
of the microwave cavity.

\section{Periodic Frequency Modulations\label{PFM}}

For periodic modulations in the frequency one must examine\cite{Wilhelm:2003}
the differential equation
\begin{eqnarray}
\ddot{\phi}(t)+\Omega^2 (t)\phi (t)=0,
\nonumber \\
\Omega^2 (t)=\Omega_0^2+\nu^2(t),
\nonumber \\
\nu(t+\tau)=\nu (t).
\label{PFM1}
\end{eqnarray}
From a mathematical viewpoint, Eq.(\ref{PFM1}) has been well
studied. If \begin{math} \nu(t) \end{math} can be represented as
a non-overlapping pulse sequence of the form
\begin{equation}
\nu (t)=\sum_{n=-\infty}^\infty \varpi(t-n\tau ),
\label{PFM2}
\end{equation}
then the transmission problem for a single pulse,
\begin{equation}
\ddot{\phi}_1(t)+\{\Omega_0^2+ \varpi^2(t)\}\phi_1 (t)=0,
\label{PMF3}
\end{equation}
yields a complete solution to the general problem. In
particular we examine the two photon creation problem as in
Eq.(\ref{DCE3}); i.e.
\begin{eqnarray}
\phi_1(t\to \infty )=e^{i\Omega_0 t}+\rho_1 e^{-i\Omega_0 t},
\nonumber \\
\phi_1 (t\to -\infty )=\sigma_1 e^{i\Omega_0 t},
\nonumber \\
|\rho_1 |^2+|\sigma_1 |^2=R_1+P_1=1,
\nonumber \\
\sigma_1=\sqrt{P_1}\ e^{-i\Theta_1}.
\label{PMF4}
\end{eqnarray}
Employing the characteristic function
\begin{equation}
\mu (\Omega_0)=\frac{\cos(\Omega_0 \tau +\Theta_1(\Omega_0))}
{\sqrt{P_1(\Omega_0)}},
\label{PMF5}
\end{equation}
one may study the stability problem for the dynamic Casimir effect.
For {\em periodic} frequency modulations there are two cases of interest:
\par \noindent
Case I: {\em Stable Motions \begin{math} -1< \mu (\Omega_0)<+1 \end{math}}
\begin{eqnarray}
\mu(\Omega_0)=\cos(\Omega \tau )
\nonumber \\
\phi_\pm (t+\tau )=e^{\pm i\Omega t}\phi_\pm (t).
\label{PMF6}
\end{eqnarray}
Case II: {\em Unstable Motions
\begin{math} \mu (\Omega_0)>+1 \end{math} {\rm or}
\begin{math}\mu (\Omega_0)<-1 \end{math}}
\begin{eqnarray}
\mu(\Omega_0)=\cosh(\gamma \tau )
\ \ {\rm or}\ \ \mu(\Omega_0)=-\cosh(\gamma \tau )
\nonumber \\
\phi_\pm (t+\tau )=e^{\pm \gamma t}\phi_\pm (t).
\label{PMF7}
\end{eqnarray}
In the unstable regime, \begin{math} 2\gamma \end{math} represents
the number of cavity photons being produced per unit time. If
the cavity mode has a high quality factor \begin{math} Q\gg 1 \end{math},
then photons are also absorbed at a rate
\begin{math} (\Omega_0/Q) \end{math}. The net photon production
rate in this approximation would then be
\begin{equation}
\Gamma_1\simeq \left(2\gamma -\frac{\Omega_0}{Q}\right),
\label{PMF8}
\end{equation}
and the theoretical noise temperature after \begin{math} n_p \end{math}
pulses would be
\begin{equation}
k_BT_1^* \approx \hbar \Omega_0 \exp(n_p \tau \Gamma_1 ).
\label{PMF9}
\end{equation}
As an example, let us suppose a sequence of rectangular pulse sequences
of the form
\begin{eqnarray}
\Omega (t)=\Omega_0 \ \ \ {\rm if}
\ \ \ t_0+n\tau < t < t_0+(n+1/2)\tau ,
\nonumber \\
\Omega (t)=(1+\alpha )\Omega_0 \ \ \ {\rm if}
\ \ \ t_0+(n+1/2)\tau < t < t_0+(n+1)\tau ,
\label{PMF10}
\end{eqnarray}
wherein \begin{math} n=1,2,\ldots ,n_p \end{math}. The estimate
\begin{equation}
\exp(n_p \tau \Gamma_1 )\sim \exp(n_p\alpha /2)
\ \ \ {\rm for} \ \ \ 1\gg \alpha \gg (\Omega_0\tau)/Q
\label{PMF11}
\end{equation}
is not unreasonable.

The exponential temperature {\em instability} for high quality
cavity modes, i.e. \begin{math} \Gamma_1 > 0 \end{math} in
Eqs.(\ref{PMF8}) - (\ref{PMF11}), would be sufficient for large
\begin{math} n_p \end{math} to {\it melt} the cavity. No microwave
oven works that efficiently even if the dynamic Casimir effect were employed
for exactly that purpose. The one loop photon approximation is evidently at
fault and higher loops (non-linear processes) must be invoked for the noise
temperature of the mode to be theoretically stable as would be laboratory
microwave cavities.

\section{Microwave Intensity Stability\label{MS}}

The stability of the microwave cavity is due to the fact that the modulation
is induced by a {\em pump} which supplies the energy of the induced cavity
radiation. One may define a {\em pump coordinate} \begin{math} \eta \end{math} which
in general is a quantum mechanical operator. In principle, one might mechanically
vibrate a wall in the cavity in which case \begin{math} \eta \end{math} would be
proportional to a mechanical displacement. In practice, changing the frequency
by electronic means may well be more efficient. Be that as it may, let us define
the coordinate so that
\begin{equation}
\left<\eta (t)\right>=\frac{\nu^2(t)}{\Omega_0^2}\ ,
\label{MS1}
\end{equation}
wherein the quantities on the right hand side of Eq.(\ref{MS1}) are given
in Eq.(\ref{PFM1}).

If the quantum pump coordinate exhibits stationary fluctuations
\begin{equation}
\Delta \eta =\eta -\left<\eta \right>
\label{MS2}
\end{equation}
with quantum noise
\begin{equation}
\frac{1}{2}\left< \Delta \eta (t) \Delta \eta (t^\prime )+
\Delta \eta (t^\prime )\Delta \eta (t)\right>=
\int_{-\infty}^\infty \bar{S}_\eta (\omega )
e^{-i\omega (t-t^\prime )}d\omega ,
\label{MS3}
\end{equation}
then two photon absorption and two photon emission processes are described
by the additional noise Hamiltonian
\begin{equation}
\Delta H=\frac{1}{4}\hbar \Omega_0
\left(a^\dagger a^\dagger+a a\right)\Delta \eta .
\label{MS4}
\end{equation}
The usual mode photon creation and destruction operators
are \begin{math} a^\dagger \end{math} and
\begin{math} a \end{math}, respectively.
When the Hamiltonian in Eq.(\ref{MS4}) is taken to second order in
perturbation theory, the resulting energies involve four boson processes
and thereby introduces multi-photon loop processes.

With the pump coordinate positive and negative frequency spectral functions
\begin{eqnarray}
\left< \Delta \eta (t) \Delta \eta (t^\prime )\right>
=\int_{-\infty }^\infty  S_\eta ^+ (\omega )
e^{-i\omega (t-t^\prime )}d\omega ,
\nonumber \\
\left<\Delta \eta (t^\prime )\Delta \eta (t) \right>
=\int_{-\infty }^\infty  S_\eta ^+ (\omega )
e^{-i\omega (t-t^\prime )}d\omega ,
\label{MS5}
\end{eqnarray}
the two photon Fermi golden rule transition rates which follow from
Eqs.(\ref{MS4}) and (\ref{MS5}) read
\begin{eqnarray}
\Gamma^+ (n\to n-2) = \frac{\pi \Omega_0^2}{8}
S_\eta ^+ (\omega =2\Omega_0)n(n-1),
\nonumber \\
\Gamma^- (n-2\to n) = \frac{\pi \Omega_0^2}{8}
S_\eta ^- (\omega =2\Omega_0)n(n-1).
\label{MS6}
\end{eqnarray}
The pump coordinate also has a noise temperature
\begin{math} T_\eta \end{math} may be defined via
\begin{equation}
S_\eta ^- (2\Omega_0)=
e^{-2\hbar \Omega_0/k_BT_\eta }S_\eta ^+ (2\Omega_0).
\label{MS7}
\end{equation}
If there a many photons in the mode, then the net
rate of photon absorption is given by
\begin{equation}
\Gamma_{absorption} \simeq 
\left(\frac{\pi \Omega_0^2\bar{S}_\eta (\omega = 2\Omega_0) }{2}\right)
\tanh\left(\frac{\hbar \Omega_0}{k_BT_\eta }\right)n^2 .
\label{MS8}
\end{equation}
On the other hand the frequency modulation produces photons
at a rate
\begin{equation}
\Gamma_{emmision} \simeq 2\gamma n
\ \ \ {\rm where} \ \ \ n \gg 1\ ,
\label{MS9}
\end{equation}
and \begin{math} \gamma  \end{math} is defined in Eq.(\ref{PMF7}). We may
now state the central result of this section:
\medskip
\par \noindent
{\bf Theorem 2:} {\it If the pump coordinate pushes the cavity mode into
a modulation dynamic Casimir instability, then the quantum noise will
stabilize the cavity mode according to the equation}
\begin{eqnarray}
\frac{dn}{dt} = 2(\gamma n - \tilde{\gamma } n^2),
\nonumber \\
\tilde{\gamma} ={\pi \Omega_0^2\bar{S}_\eta (\omega = 2\Omega_0) }
\tanh \left(\frac{\hbar \Omega_0}{k_BT_\eta }\right).
\label{MS10}
\end{eqnarray}
\medskip
\par \noindent
The cavity photon occupation number will then saturate according to
\begin{equation}
\bar{n}_{saturate}=
\frac{\gamma }{\pi \Omega_0^2 \bar{S}_\eta (\omega = 2\Omega_0) }
\coth \left(\frac{\hbar \Omega_0}{k_BT_\eta }\right).
\label{MS11}
\end{equation}
More simply, with the response function 
\begin{equation}
\chi(\zeta )=\frac{i}{\hbar}\int_0^\infty 
\left<\left[\eta (t),\eta (0)\right]\right>e^{i\zeta t}dt,
\label{MS12}
\end{equation}
the fluctuation dissipation theorem 
\begin{equation}
\bar{S}_\eta (\omega)=\left(\frac{\hbar }{2\pi}\right)
\coth\left(\frac{\hbar \omega }{2k_BT_\eta }\right){\Im m}\chi(\omega +i0^+).
\label{MS13}
\end{equation}
together with Eqs.(\ref{MS11}) 
and (\ref{MS12}) reads
\begin{equation}
\bar{n}_{saturate}=
\frac{2\gamma }{ \Omega_0^2 [\hbar  {\Im m}\chi(2\Omega_0 +i0^+)]}.
\label{MS14}
\end{equation}
The relation time \begin{math} \tau^\dagger \end{math} for the parameter 
\begin{math} \eta \end{math} may be conventionally 
defined\cite{Martin:1968} by 
\begin{equation}
\chi(0)\tau^\dagger = 
\lim_{\omega \to 0}\frac{{\Im} m\chi(\omega +i0^+)}{\omega }
\label{MS15}
\end{equation}
so that 
\begin{equation}
\bar{n}_{saturate}\approx 
\frac{\gamma }{ \Omega_0^3 \tau^\dagger \hbar \chi(0)}.
\label{MS16}
\end{equation}
Eq.(\ref{MS16}) is our final answer for the number of final photons 
at saturation.

\section{A Numerical Example \label{ANE}}

In order to make our final answer less abstract, let us consider a 
proposed\cite{Braggio:2004} experiment. In that proposal, the parameter 
\begin{math} \eta  \end{math} describes the metallic conductivity in a 
semiconductor plate due to a laser beam inducing particle hole pairs. 
If we let \begin{math} \tau_R  \end{math} represent the recombination 
time taken to annihilate a particle hole pair in the semiconductor 
and let \begin{math} \omega_L  \end{math} represent the laser frequency, 
then we estimate that 
\begin{equation}
\frac{1}{\tau^\dagger}\sim\frac{\hbar \omega_L\chi(0)}{\tau_R}
\label{ANE1}
\end{equation}
which implies
\begin{equation}
\bar{n}_{saturate}\sim
\left(\frac{\gamma }{ \Omega_0}\right)
\left(\frac{1}{\Omega_0 \tau_R}\right)\left(\frac{\omega_L}{\Omega_0}\right).
\label{ANE2}
\end{equation}
The following estimates are reasonable for the proposal\cite{Braggio:2004}:
\begin{eqnarray}
\left(\frac{\gamma }{ \Omega_0}\right)\sim 0.05,
\nonumber \\
\left(\frac{1}{\Omega_0 \tau_R}\right)\sim 10,
\nonumber \\
\left(\frac{\omega_L}{\Omega_0}\right)\sim 2\times 10^5,
\nonumber \\
\bar{n}_{saturate}\sim 10^5 \ {\rm microwave\ photons.}
\label{ANE3}
\end{eqnarray}

\section{Conclusion\label{Conc}}

We have explored the concept of induced instabilities in both the static
and dynamic Casimir effects. For the static case, large quantum
electrodynamic collective Lamb shifts in condensed matter can
induce a phase transition requiring a new equilibrium position
of the microwave oscillator coordinates. In particular, when at the quadratic
level and oscillator goes unstable, quartic terms can be invoked to make the
system stable. For the dynamic case, even if the frequency shifts are small,
perfect periodicity in modulation pulses can build up to exponentially large
proportions again leading to an instability. Again dynamic quartic terms
can stabilize the cavity modes. The basic principle involved is that
the shifted frequencies themselves must undergo fluctuations. Given the noise
fluctuations in the pump coordinate, the final saturation temperature
of the microwave cavity can be computed from Eq.(\ref{MS16}).


\vskip 0.5cm

\begin{thebibliography}{04}

\bibitem{Casimir:1948}
H.B.G. Casimir and D. Polder, {\it Phys. Rev.}
{\bf 73}, 360 (1948).

\bibitem{Casimir::1948}
H.B.G. Casimir {\it Proc. K. Ned. Akad. Wet.} {\bf 60}, 793 (1948).

\bibitem{Bordag:2001}
M. Bordag, U. Mohideen, and V. M. Mostepanenko,
{\it Phys. Rep.} {\bf 205}, 353 (2001).

\bibitem{Milton:2001}
K. A. Milton, {\it ``The Casimir Effect''}, World Scientific,
River Edge, (2001).

\bibitem{Dzyaloshinski:1960}
I.E. Dzyaloshinski, E.M. Lifshitz and L.P. Pitayevski,
{\it Sov. Phys. JETP} {\bf 10}, 161 (1960).

\bibitem{Dzyaloshinski:1961}
I.E. Dzyaloshinski, E.M. Lifshitz and L.P. Pitayevski,
{\it Adv. Phys.} {\bf 10}, 165 (1961).

\bibitem{Dalvit:1999}
D.A.R. Dalvit and F.D. Mazzitelli,
{\it Phys. Rev.} {\bf A 59}, 3049 (1999).

\bibitem{Dodonov:1998}
V. V. Dodonov, {\it J. Phys.} {\bf A 31} 9835, (1998).

\bibitem{Dodonov:1999}
V. V. Dodonov and M. A. Andreata,
{\it J. Phys.} {\bf A 32} 6711 (1999).

\bibitem{Widom:1987}
Y. Srivastava and A. Widom.
{\it Phys. Rep.} {\bf 148}, 1 (1987).

\bibitem{Widom:2004}
S. Sivasubramanian, A. Widom and Y.N. Srivastava,
arXiv:cond-mat/0403592, (2004).

\bibitem{Wilhelm:2003}
W. Magnus and S. Winkler, {\it ``Hill's Equation''},
Dover Publications (2003).

\bibitem{Yablonovitch:1989}
E. Yablonovitch, {\it Phys. Rev. Lett.} {\bf 62}, 1742 (1989).

\bibitem{Braggio:2004}
C. Braggio, G. Bressi, G. Carugno, C. Del Noce, G. Galeazzi, A. Lombardi,
A. Palmieri, G. Ruoso and D. Zanello, arXiv:quant-ph/0411085 (2004).

\bibitem{Martin:1968}
P.C. Martin, {\it ``Measurements and Correlation Functions''}, 
Gordon and Breach Scientific Publishers, New York (1968).

\end{thebibliography}
\end{document}